\documentclass[conference]{IEEEtran}
\IEEEoverridecommandlockouts
\usepackage{cite}
\usepackage{amsmath,amssymb,amsfonts}
\usepackage{algorithmic}
\usepackage{graphicx}
\usepackage{textcomp}
\usepackage{xcolor}
\usepackage{lineno}
\usepackage{booktabs} 
\usepackage{subfig}
\usepackage{lineno,hyperref}
\usepackage{multirow}
\usepackage{multicol}
\modulolinenumbers[5]
\usepackage{csquotes}
\usepackage{booktabs} 
\usepackage{braket}
\usepackage{multirow}
\usepackage{multicol}
\usepackage{graphicx}
\usepackage{amssymb}
\usepackage{amsmath}
\usepackage{url}
\usepackage{rotating}
\usepackage{color}
\usepackage{makecell}
\usepackage[framemethod=TikZ]{mdframed}
\usepackage{graphicx}
\usepackage{subfig}
\usepackage{xcolor}
\usepackage{tcolorbox}
\def\BibTeX{{\rm B\kern-.05em{\sc i\kern-.025em b}\kern-.08em
    T\kern-.1667em\lower.7ex\hbox{E}\kern-.125emX}}

\begin{document}

\title{Quantum Computing Platforms: Assessing the Impact on Quality Attributes and SDLC Activities\\

}
\author{Anonymous}
\author{\IEEEauthorblockN{ Balwinder Sodhi}
\IEEEauthorblockA{\textit{Department of Computer Science and Engineering} \\
\textit{Indian Institute of Technology}\\
Ropar, Punjab, India. \\
sodhi@iitrpr.ac.in}
\and
\IEEEauthorblockN{ Ritu Kapur}
\IEEEauthorblockA{\textit{Department of Computer Science and Engineering} \\
\textit{Indian Institute of Technology}\\
Ropar, Punjab, India. \\
dev.ritu.kapur@gmail.com}
}

\maketitle

\begin{abstract}
Practical quantum computing is rapidly becoming a reality. To harness quantum computers' real potential in software applications, one needs to have an in-depth understanding of all such characteristics of quantum computing platforms (QCPs), relevant from the Software Engineering (SE) perspective. Restrictions on copying, deletion, the transmission of qubit states, a hard dependency on quantum algorithms are few, out of many, examples of QCP characteristics that have significant implications for building quantum software. 

Thus, developing quantum software requires a paradigm shift in thinking by software engineers. This paper presents the key findings from the SE perspective, resulting from an in-depth examination of state-of-the-art QCPs available today. The main contributions that we present include \textbf{i)} Proposing a general architecture of the QCPs, \textbf{ii)} Proposing a programming model for developing quantum software, \textbf{iii)} Determining architecturally significant characteristics of QCPs, and \textbf{iv)} Determining the impact of these characteristics on various Quality Attributes (QAs) and Software Development Life Cycle (SDLC) activities.

We show that the nature of QCPs makes them useful mainly in specialized application areas such as scientific computing. Except for performance and scalability,  most of the other QAs (e.g.,  maintainability,  testability,  and reliability) are adversely affected by different characteristics of a QCP.
\end{abstract}


\section{Introduction}
\label{sec:intro}
The origin of quantum information processing can be traced back to the early 1970's, however, it was only in 2011 that the first commercially viable quantum computer was reported  \cite{dwave-opens-box-2011}. Few recent QCP space developments include Intel's delivery  \cite{intel-qpu-2017} of a 17-qubit superconducting quantum chip and the IBMs  \cite{ibm50qbit2017} 50-qubit quantum computer. Today, it is possible for a programmer to use a real quantum computer through cloud-based quantum programming platforms (e.g., IBM Q Experience  \cite{ibm-qx}).

From the perspective of software development, \textit{quantum computing} is one of the most recent paradigm shifts. In this paper, we use the term \textit{Quantum Computing Platform (QCP)} to refer to the entire apparatus (hardware and software), which is necessary to develop and deploy \textit{quantum software} applications. A typical software application that runs on a QCP consists of some \textit{native quantum} parts and some classical, non-quantum software parts. As of today, the native quantum parts are of a relatively small size. Considering the pace of development happening in the domain of QCPs, it is pertinent to investigate how the large-scale professional development of quantum software be different (or similar) to building software for purely classical computing platforms. Some of the critical questions in this context could be:
\begin{enumerate}
\item What are the essential characteristics of QCPs, which are efficacious for software development?
\item In what way does a QCP affect the quality attributes (and non-functional requirements) of quantum software applications?
\item Which Software Development Life Cycle (SDLC) activities are affected, and how, by characteristics of a QCP?
\end{enumerate}

In this paper, we address these questions via an in-depth examination of the QCPs available today. 

\begin{figure}[h]
      \centering
      \includegraphics[width=\linewidth]{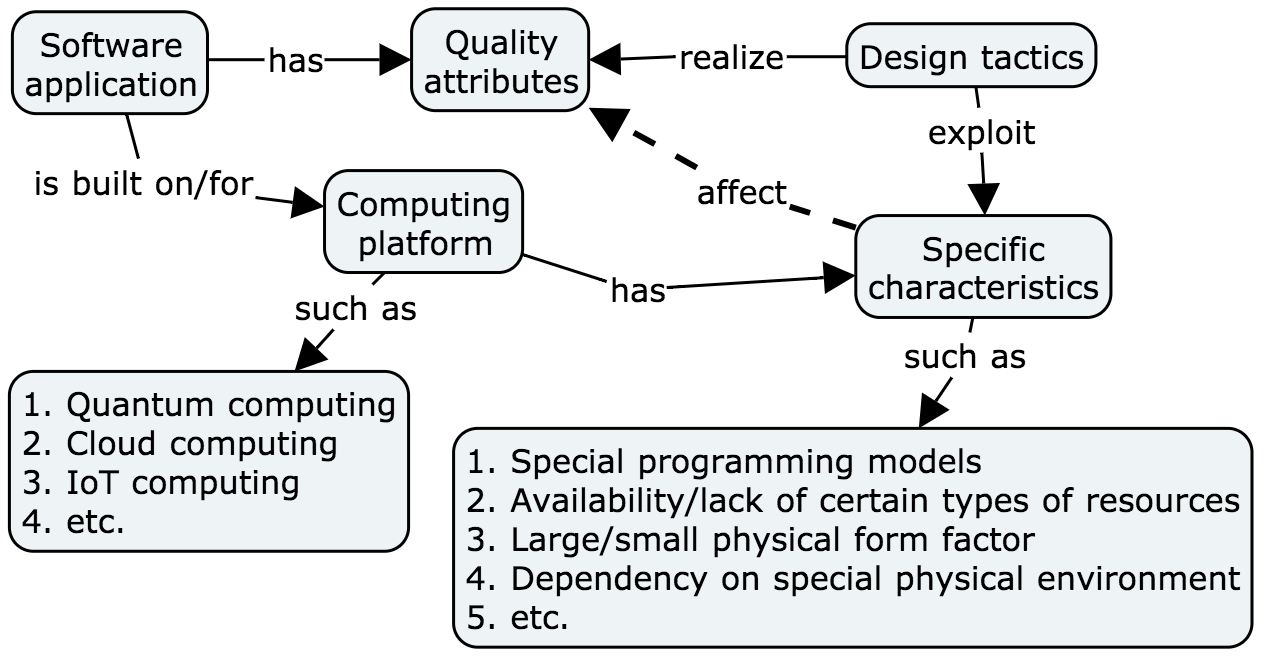}
      \caption{Relationship between QAs and computing platform characteristics.}
      \label{fig:qa-chars}
   \end{figure}

\subsection{Method of research}
\label{sec:method}
Our work's central tenet is that \textit{a computing platform's characteristics affect design and software development for that platform.} In the past, the authors \cite{sodhi-assess-platform-qa-2012,Sodhi:2011-assess-cloud} have established from the experimental findings that a computing platform's characteristics affect the QAs of a software application. Such a relationship between QAs and platform characteristics is expected because QAs are realized  \cite[Part-II]{bass-sap-3ed-2012} using design tactics, and the tactics leverage  \cite{bachmann2002illuminating,deriving-tactics,bass2000quality} characteristics of the computing platform, illustrated in Fig. \ref{fig:qa-chars}. Because the items such as QAs and tactics always occur in the context of software development, the above results imply that the SDLC activities are also affected by a computing platform's characteristics.

To understand the above point, consider this example: building the \textit{performance} QA into an application often demands to exploit parallelism present in a task. However, if the target computing platform does not offer multiprocessing (e.g., if it has only one single-core CPU), the application will not truly realize the QA of performance.

Our approach for the present work, which seeks to find answers to questions highlighted in the preceding section, is based on the above findings. The central idea underlying our approach can be stated as follows:
\begin{enumerate}
\item Identify a set, $\Theta$, of crucial characteristics of QCPs by an in-depth examination of the available literature and software artifacts on various QCPs.
\item Validate each $\theta \in \Theta$ via prototyping or reference data as necessary.
\item Identify the cause-and-effect relationships that each $\theta \in \Theta$ has with the QAs  \cite{bachmann2002illuminating} and SDLC activities  \cite{sw-maint-economics-2006}. This step is performed by examining the definition for each QA  \cite[Part-II]{bass-sap-3ed-2012} and SDLC activities  \cite{swebokv3}.
\end{enumerate}

These steps were carried out by independent professional software engineering experts who had experience in building complex software. The Delphi method  \cite{delphi-book-murray-1975} was used to arrive at the consensus about mixed results. This approach's application is demonstrated in Section \ref{sec:qcp-char-impact}, where we determine QCP characteristics' impact on various QAs and SDLC activities. However, we do not develop any quantum software applications while conducting our research. The study is purely an experience report based on \enquote{experimental programming} by professional software engineering experts.

The paper's structure broadly reflects how we carried out this work: We start by describing in Section \ref{sec:q-platforms} the quantum computing basics relevant from the perspective of developing quantum software. By in-depth examination of different QCPs available from various providers, we bring out in Section \ref{sec:a-platform-arch} the general architecture of a QCP. In Section \ref{sec:q-software-dev}, we present the techniques exploited by currently known quantum algorithms and propose a programming model for quantum software development. In Section \ref{sec:implementation details}, we present the details of experiments performed with various quantum computing toolkits to determine the pros and cons in working at various QCPs. In Section \ref{sec:identify-qcp-char-and-impact}, we determine the prominent QCP characteristics, and identify the QAs and SDLC activities affected by them. 
In Section \ref{sec:qcp-chars-and-qa-impact}, we present the characteristics of QCPs and how they impact various QAs. Finally, we discuss the effect of these characteristics on SDLC activities in Section \ref{sec:effect-swp}.

\section{Quantum computing platforms}
\label{sec:q-platforms}
Determining software architectural aspects of quantum software development requires deeper comprehension of essential differences between the classical and the quantum computers. Though a detailed treatment of the theoretical underpinning of the quantum phenomenon and computing is available in standard quantum computing texts such as  \cite{nielsen2002quantum}, we present quantum computing concepts necessary for our discussion. The description of these concepts is necessary to set a ground for the identification of the necessary quantum computing characteristics and study their impact on various SDLC activities and QAs.

\subsection{What makes quantum computing different?}
\label{sec:key-quantum-principles}
A classical computer stores and processes information in the form of binary bits using 1 and 0. A bit is realized at a physical level using suitable properties (e.g., voltage or current) of some physical device.  An $n$ bit memory cell in a classical computer can potentially represent $2^n$ different \emph{symbols}, but at any moment, it can represent \emph{only one} of these $2^n$ possibilities. Information on a quantum computer is represented or stored using quantum bits (or qubits). These qubits are physically realized via suitable physical phenomena that obey the laws of quantum mechanics. 
Examples of such phenomena can be \enquote{The spin of a single electron or an individual ion's configuration.} A qubit can be \textit{observed} in one of the two \textit{basis states}, labeled as $\ket{1}$ or a $\ket{0}$. Generally, the qubit state $\ket{1}$ corresponds to classical bit 1, and $\ket{0}$ to classical bit 0. One of the remarkable properties of qubits, which allows quantum computers to compute much faster than classical computers, is the number of possible states in which a qubit can \textit{exist} (different from its observable state, which can be only one of the basis states). At any instant, one qubit can be in a $\ket{1}$, a $\ket{0}$ state, or any \textit{quantum superposition}  \cite{nielsen2002quantum} of the two. In other words, the basis states $\ket{1}$ and $\ket{0}$, and their linear combinations $x \ket{1} + y \ket{0}$ describe all the possible states of a single qubit that can exist at any given time. In contrast, one classical bit at one time can be \emph{only in one} of the two possible states: a 0 or 1.

\begin{tcolorbox}[boxrule=1pt]
\textbf{The relevance of this property for computing:}
A quantum computer can represent a much larger state space with the same number of qubits as the classical computer's bits, which is because a qubit $\psi = x \ket{1} + y \ket{0}$ can be thought of as a vector in a complex plane. The \enquote{phase difference} between $\ket{1}$ and $\ket{0}$ basis states of the qubit represents a significant quantity. Certain types of probabilistic algorithms can exploit this property to perform faster computations.
\end{tcolorbox}

Another quantum property that is exploited by quantum computing is the \textit{entanglement} of qubits. Two or more individually independent quantum objects are said to be entangled when a) their behavior is random individually, but at the same time, b) it is strongly correlated despite each object being independent of the other. A multi-qubit state that cannot be expressed as a list of the individual constituent qubits is entangled. Consider the Bell State  \cite{bellstate} $q_1q_2 := (\ket{00} + \ket{11})/\sqrt{2}$. It is an example of an entangled two-qubit state. There is no way of expressing it as a list of one-qubit states. Suppose we measure (at some axis) one of the qubits, say $q_1$, that comprise this entangled state. It will behave randomly: in this case, $q_1$ can be $\ket{0}$ or $\ket{1}$ with equal probability. Suppose we measured $q_1$ to be $\ket{0}$, then the value of $q_2$ will certainly be $\ket{0}$. That is, we can predict exactly how the other qubit, $q_2$, would behave if measured along the same axis. No unentangled state exhibits this type of perfect correlation with perfect individual randomness.

\begin{tcolorbox}[boxrule=1pt]
\textbf{The relevance of this property for computing:}
A set of independent random variables or phenomena can be modeled using a set of qubits. If some of these qubits are chosen to be entangled, it is possible to develop algorithms that need to operate only on a subset of these entangled qubits, thus making the computation faster. There could be many other scenarios in which the property of entanglement can be useful.
\end{tcolorbox}

The above two properties -- \textit{quantum superposition} and \textit{quantum entanglement} -- are useful resources in quantum computing. The following \emph{no-go theorems} put additional constraints on how computation may be carried on a QCP:

The \textbf{no-cloning theorem  \cite{no-cloning-wootters1982}} states that \textit{it is impossible to create an identical copy (or clone) a quantum state}, which is because it is impossible to find the actual qubit state at any point in time. Therefore, it is  not possible to clone an unknown quantum state.

The \textbf{no-communication theorem \cite{popescu1998causality}}  states that \textit{during measurement of an entangled quantum state, it is not possible to make a measurement of a subsystem of the total state, and to communicate the same.} The theorem disallows all kinds of communication through shared quantum states.


The \textbf{no-teleportation theorem \cite{pathak2013elements}} states that \textit{an arbitrary quantum state cannot be converted into a sequence of classical bits.} It also states the reverse to be false.

The \textbf{no-deleting theorem  \cite{pati2000impossibility}} states that, \textit{ given two copies of an arbitrary quantum state, it is impossible to delete one of them.}

  \begin{figure}[h]
      \centering
      \includegraphics[width=0.8\linewidth]{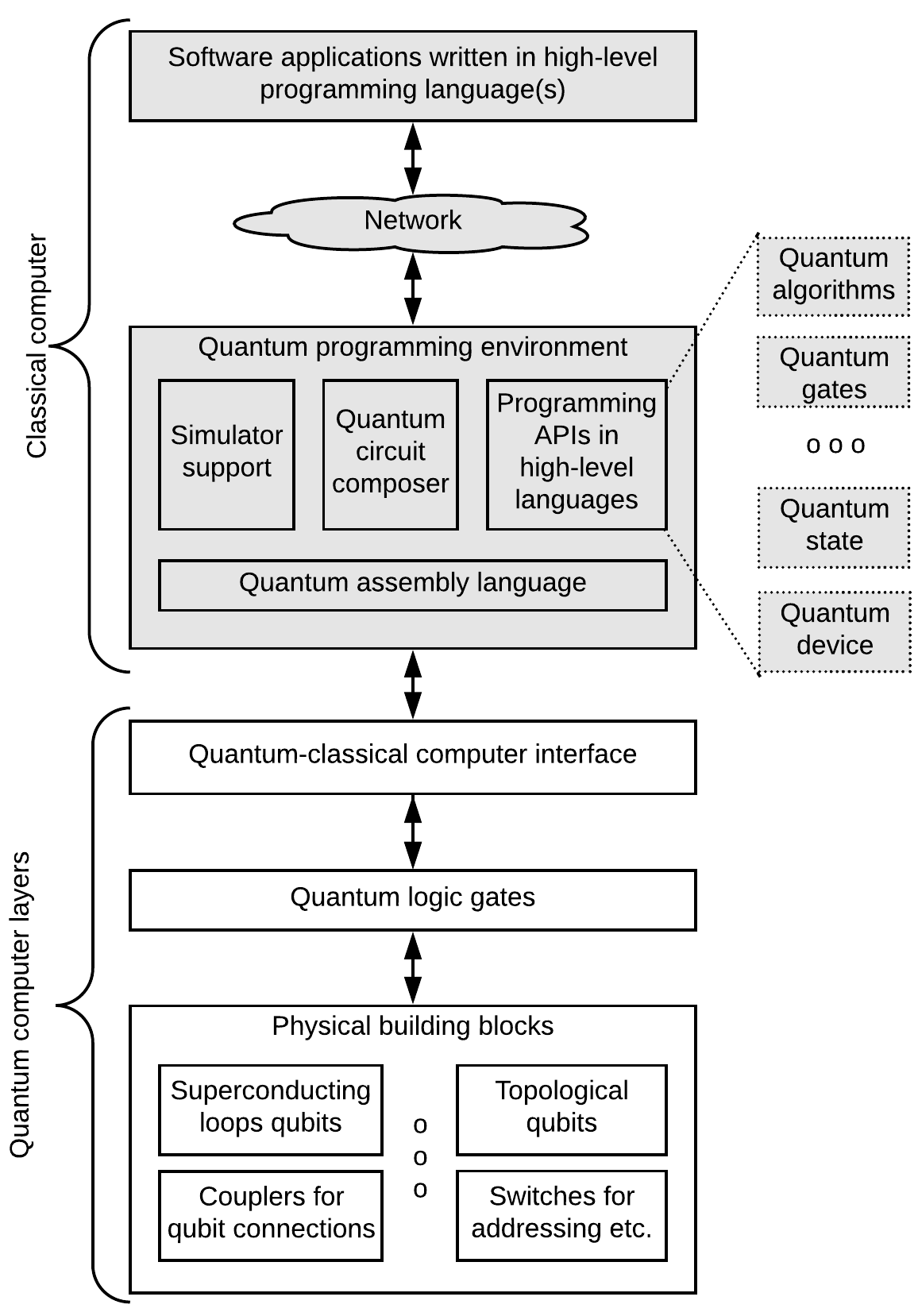}
      \caption{Architecture of quantum computing platform.}
      \label{fig:arch}
   \end{figure}
   
 \subsection{Related Work}
 Authors in \cite{fingerhuth2018open} present a study of 24 Open Source Quantum computing tools covering various details of quantum toolchain and different quantum computing paradigms. The authors study the characteristics such as the documentation, licence, the choice of programming language, compliance with norms of software engineering used in these tools. Authors report that it is a challenge to select a suitable paradigm and the corresponding quantum algorithm to solve a specific problem, which supports one of our claims too. The study provides a good insight of the features present in various toolkits. Based on their study the authors conclude that it is hard for the newcomers to become handy with the quantum computing toolkits, and is difficult for experienced quantum developers to switch-to the open-source toolkits.
 
 Authors in \cite{larose2019overview} provide a comparison of four cloud-based Quantum computing platforms: Rigetti's Forest (pyQuil), IBM's Qiskit, ProjectQ, and Microsoft's Quantum Developer Kit (Q\#). We also experimented with three of these toolkits with the objective of gaining an experience of their inherent features and the support offered for a developer. The authors provide the details related to various aspects of these toolkits, such as their  requirements and installation, language syntax, library support, and quantum simulator capabilities. The intent of the authors lies in providing a feature comparison of various quantum toolkits, a quick installation guide, and some examples to guide a beginner for a quick start with the toolkits.
 
Authors in \cite{leymann2019towards} introduce a novel platform for sharing quantum software (QuSP) that connects various software community members such as the quantum-domain experts, developers, and end-users to work on the existing quantum computing algorithms available in different sources of information, such as books, web, and research papers. The work is currently in progress towards developing a prototype for QuSP. 

Authors in \cite{weder2020quantum} introduce a quantum-specific SDLC titled as Quantum Software Lifecycle (QSLC) for developing a gate-based quantum software application. The QSLC comprises ten software development phases, viz., quantum classical splitting, hardware-independent implementation, quantum circuit enrichment, hardware-independent optimization, quantum hardware selection, readout-error mitigation preparation, compilation and hardware-dependent optimization, integration, execution, and results and analysis. The authors provide a comprehensive description of the goals, methods, applicable tools, and viable input/output data for the QSLC phases.      

Authors in \cite{zhao2020quantum} provide a survey of the various quantum software engineering techniques. The authors categorize the existing literature as per the quantum techniques present for various SDLC activities. The review provides the details of a variety of quantum software engineering definitions, quantum concepts, and how quantum computing differs from classical computing. The authors point out the negligence towards quantum software reuse as a critical issue. 

Authors in \cite{leymann2020bitter} discuss the challenges in implementing the gate-based Quantum Computing Algorithms on a noisy intermediate scale quantum (NISQ) device. The authors discuss the essential concepts related to the state preparation, oracle expansion, connectivity, circuit rewriting, and readout, which play a significant role in implementing the quantum algorithms. Authors validate their findings by performing several experiments on the IBM quantum computing toolkit, and provide a detailed discussion.

Clearly, there exist many useful sources of gaining an insight on quantum computing basics and quantum toolkits. But, to the best of our knowledge, there does not exist any work that studies the impact of quantum computing characteristics on SDLC activities and QAs.   

\subsection{General architecture}
\label{sec:a-platform-arch}
Today's physical quantum computing machinery (reminiscent of early classical computers of the 1940s) is significant  \cite{dwave-hardware}. It requires a particular physical environment and conditions for operating correctly. We examined the QCPs available today from various providers (e.g., \cite{dwave-opens-box-2011,ms-quantum,ibm-qx,rigetti-qpu}). Considering the salient characteristics of such QCPs, we propose the general architecture of a QCP, as depicted in Fig. \ref{fig:arch}. It comprises of five layers, three of which contain purely quantum hardware and circuitry and two consists of classical hardware and software:
\begin{enumerate}
  \item \textbf{Quantum layers}  \cite{dwave-hardware,quantum-von-nueman-arch,ms-quantum}
  One can think of these layers as comprising the Quantum Processing Unit (QPU).
    \begin{enumerate}
        \item \textit{Physical building blocks --} Includes quantum hardware that typically uses superconducting loops for physical realization of qubits. It also contains the physical qubit coupler or the interconnect-circuitry, among other elements needed for qubit addressing and control operations.
        \item \textit{Quantum logic gates --} Physical circuitry  \cite[\S 5.5]{quantum-von-nueman-arch} that makes up quantum logic gates \cite{nielsen2002quantum}.
        \item \textit{Quantum-classical interface --}  Includes the hardware and software that provides interfacing between classical computers and a QPU.
    \end{enumerate}
  \item \textbf{Classical layers} \cite{rigetti-qpu,dwave-tech-overview,ibm-qx,ms-quantum-sdk}
    \begin{enumerate}
        \item \textit{Quantum programming environment --} It provides items such as i) quantum assembly language necessary for instructing a QPU, ii) the programming abstractions needed for writing quantum programs in a high-level programming language, and iii) simulator support and IDEs.
        \item \textit{Software applications --} Quantum software applications are written to cater for business and scientific requirements.
    \end{enumerate}
\end{enumerate}

\subsection{Development of quantum software}
\label{sec:q-software-dev}
In the preceding section, we introduced the two fundamental properties -- \textit{quantum superposition} and \textit{quantum entanglement} -- which make a quantum computer much faster than classical computers at solving specific problems. The nature of these two properties inherently makes the quantum computation to be mostly probabilistic. Thus, implying that quantum programs are likely to be probabilistic or randomized. Expressing the logic or algorithm that is to be executed on a quantum computer requires special techniques and programming models.

\subsubsection{Quantum algorithms -- The key to harnessing power of QCPs}
\label{sec:q-problem-modeling}
The usefulness of quantum programs lies in their ability to exploit the fundamental characteristics (superposition and entanglement of qubits) of a quantum computer. Over the past few decades, quantum computing researchers have developed a handful of techniques \cite{cleve1998quantum} that exploit quantum computers' characteristics for quickly solving specific problems that take much longer to solve on a classical computer. Some of these techniques and example algorithms that exploit them are shown in Table \ref{tab:q-algo-techniques}. Several algorithms are available today which exploit these techniques to solve various computing problems much faster than a classical computer. To derive benefit from a quantum computer's capabilities, one must map the problem at hand to one of the problems for which a quantum algorithm is known.

\begin{table}[!t]
\caption{Techniques used by quantum algorithms}
\label{tab:q-algo-techniques}
\begin{center}
\resizebox{\linewidth}{!}{
\begin{tabular}{c c} 
\toprule
\textbf{Techniques} & \textbf{Examples of the target problems}\\ \midrule
\multirow{2}{*}{\shortstack{Amplitude\\ amplification  \cite{quantum-amplitude-amplifiy}}} & \multirow{2}{*}{\shortstack{Searching in a collection or database-\\ of items (Grover's algorithm \cite{grover1996fast}).}} \\ 
&\\\hline
\multirow{5}{*}{\shortstack{Quantum Fourier\\ transform  \cite{quantum-fft}}} & \multirow{5}{*}{\shortstack{Discrete logarithm problem and the integer-\\ factorization problem in polynomial time (Shor's-\\ algorithm), and so on. Predominant applications\\ are in the field of cryptography.}}\\ 
&\\
&\\
&\\
&\\ \hline
\multirow{4}{*}{\shortstack{Phase kick-\\back  \cite{cleve1998quantum}}} & \multirow{4}{*}{\shortstack{Estimating Gauss sums \cite{van2002efficient}. Application areas-\\ include: simulations of complex physical-\\ processes, fluid dynamics, etc.}} \\ &\\
&\\
&\\\hline
\multirow{5}{*}{\shortstack{Quantum phase\\ estimation  \cite{cleve1998quantum}}} & \multirow{5}{*}{\shortstack{Estimates the phase that a unitary transformation\\ adds to one of its eigen-vectors.\\ Application areas include: simulations of-\\ complex physical processes, fluid dynamics, etc.}}\\ 
&\\
&\\
&\\
&\\\hline
\multirow{5}{*}{\shortstack{Quantum\\ walks  \cite{quantum-walks-review,quantum-walks-algo-apps}}} & \multirow{5}{*}{\shortstack{Element distinctness problem, Triangle finding-\\ problem, Formula evaluation, Group commutativity.\\ Application areas include: flow analysis,\\ networks, natural language processing, etc.}} \\ 
&\\
&\\
&\\
&\\\hline
\end{tabular}
}
\end{center}
\end{table}

   \begin{figure*}
      \centering
      \includegraphics[width=0.7\linewidth]{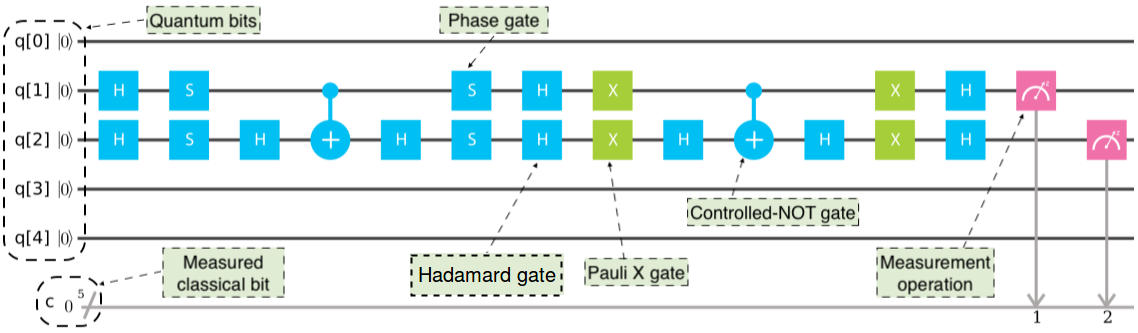}
      \caption{An example quantum circuit developed using IBM's quantum computing simulator.}
      \label{fig:composer-grover-algo}
   \end{figure*}

Quantum circuits  \cite{quantum-circuits} is one of the standard models for representing quantum computation. 
Like digital logic gates employed by classical computers, \textit{quantum gate}s are used to compose a quantum circuit. In this model, a quantum algorithm's steps can be expressed as a sequence of quantum logic gates.
Each quantum logic gate transforms the input qubits in a well-defined manner, typically expressed as operations on matrices and vectors. IBM Q-Experience Composer  \cite{ibm-qx} takes this approach to expressing quantum computations. Fig. \ref{fig:composer-grover-algo} shows the screenshot of a quantum circuit for implementing Grover's algorithm \cite{grover1996fast} on IBM Q-Experience \cite{ibm-qx}. The square boxes on horizontal lines represent the quantum gates used in this circuit. Each of the horizontal lines (except the one at the bottom) represents a qubit's lifeline. Different quantum gates can be applied to a qubit along its lifeline from left to right. A measurement operation is typically the last item on the lifeline of a qubit.

   \begin{figure}
      \centering
      \includegraphics[width=0.8\linewidth]{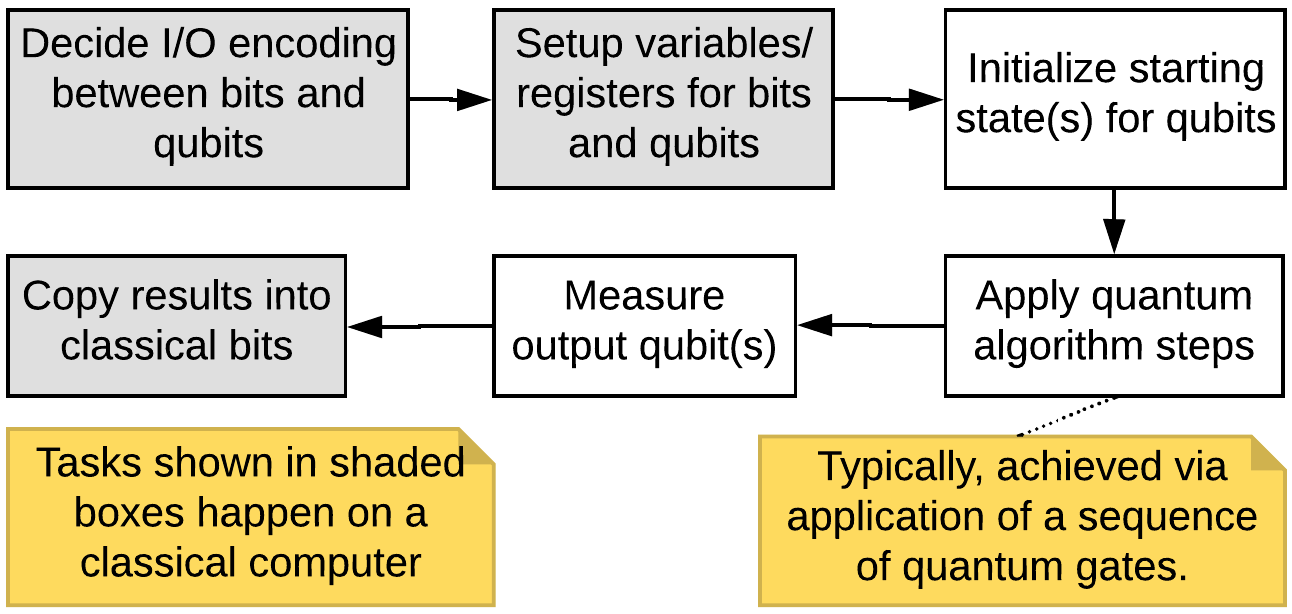}
      \caption{A quantum programming model.}
      \label{fig:q-prog-model}
   \end{figure}

\subsubsection{Programming model in quantum paradigm}
\label{sec:q-programming-model}
We evaluated the programming abstractions, the high-level programming language libraries that expose these abstractions, and the typical steps involved in composing quantum programs on different QCPs such as  \cite{ibm-qx,dwave-tech-overview,ms-quantum-sdk,rigetti-qvm}. We observe that a quantum program typically comprises parts that execute on a classical CPU and some on a Quantum Processing Unit (QPU). Creating such quantum programs mainly involved the following tasks:
\begin{enumerate}
\item Mapping input/output (I/O) from classical bits representation to qubits.
\item Initializing the qubits state.
\item Compose the quantum circuit using suitable quantum logic gates to express a quantum algorithm's steps. The steps are repeated a suitable number of times to get a reliable measurement of the outcomes.
\item Measure the output qubits state (measuring a qubit forces it to collapse into a classical bit) and transfer it to the classical bits.
\end{enumerate}

A suitable high-level programming API or instructions \cite{ms-quantum-sdk,open-qasm,rigetti-quantum-isa-2016} is typically used to compose the quantum programs. Based on these observations, a quantum programming model has been derived by us, shown in Fig. \ref{fig:q-prog-model}. We also checked the adherence to this programming model by manually examining the sample quantum programs supplied by different QCPs.
 
\subsection{Experimenting with major QCPs}
 \label{sec:implementation details}
To discover the characteristics of various QCPs, we experimented with the three major quantum programming platforms, viz., IBM's Quantum Experience platform (Composer and QASM Editor)  \cite{ibmquant}, Microsoft's Quantum Development Kit  \cite{microsoft-quant}, and Rigetti's Quantum Computing Services  \cite{rigetti}. These QCPs are freely accessible, have user-friendly interfaces.

\subsubsection{IBM's Quantum Experience Platform}
\label{sec:IBM-Quantum-Experience-Simulator}
IBM's QCP provides a user with an easy interface of quantum logic gates to build a quantum circuit. We experimented with a subset of the state-of-the-art quantum computing algorithms  \cite{coles2018quantum} using the IBM 5-qubit QCP named \texttt{ibmqx4}. For example, the circuit diagram of Grover's algorithm implemented using \texttt{ibmqx4} is shown in Fig. \ref{fig:composer-grover-algo}. The programming interface for implementing Quantum Computing algorithms works on \emph{Quantum assembler (Qasm) language} and is called the QASM editor. 

IBM provides support for Quantum Computing as a cloud service. Being a cloud service, all benefits of a cloud based platform are available here.

\subsubsection{Microsoft's Quantum Development Kit (QDK)}
\label{sec:Microsoft-Quantum-Development-Kit}
Microsoft's QDK works in the Visual Studio environment\footnote{\url{https://docs.microsoft.com/en-us/quantum/?view=qsharp-preview}} and can be installed on a machine having Windows, macOS, or a Linux operating system. Code samples for some prominent Quantum algorithms are freely available on GitHub  \cite{microsoft-quant-git-code-samples}. We implemented a subset of these algorithms (screenshots, etc. of our setup are available at \url{https://bit.ly/3j37PlX}).

Microsoft's QDK provides good documentation and developer support resources. However, the programming interface is via QASM, thus imposes additional learning curve for the developers not accustomed to low level programming languages.

\subsubsection{Rigetti's  Quantum Computing Services (QCS)}
\label{sec:Rigetti-Quantum-Computing-Services}
Rigetti's QCS\cite{rigetti-forest-sdk} consists of the \texttt{pyQuil} (Python Quantum Instruction Language)\footnote{\url{https://readthedocs.org/projects/pyquil/}} library, the Rigetti Quil Compiler (quilc), and the Quantum Virtual Machine (qvm). The quilc compiles the quantum programs written in quil to be run on a Quantum Abstract Machine (QAM). Rigetti's qvm is a QAM provided for implementing various quantum algorithms on a local machine. 
To run a Quantum script file, one needs to initialize the quilc and the qvm server (hosted locally). When a quantum file is run, the quil instructions (which are in python) are compiled into the native quil form by the quilc. The compiled code is then run on the qvm server instance to produce the results. We implemented a subset of quantum computing algorithms' snippets from \cite{rigetti-quant-git-code-samples} by using this SDK.

Rigetti's QCS allows implementing quantum algorithms in a well-known programming language -- \texttt{python}.

\section{Determining QCP characteristics and their impact}
\label{sec:identify-qcp-char-and-impact}
We now describe our approach for determining the salient characteristics of QCPs and how each of these characteristics impacts various QAs and SDLC activities.  
\subsection{Determining QCP characteristics}
\label{sec:determine-qcp-chars}
Most of the QCPs available today provide a set of sample software programs that demonstrate the capabilities of respective QCPs. These QCPs also provide the detailed documentation artifacts (e.g.  \cite{ms-quantum-sdk,open-qasm,rigetti-quantum-isa-2016}) that explain the internals of these QCPs and the high-level APIs that allow programming on those QCPs. By carefully examining them and prototyping with these artifacts, we have identified the key characteristics (presented in Section \ref{sec:qcp-char-impact}) of the QCPs. We have cited relevant artifacts of the QCPs when presenting the respective characteristics in the rest of this paper.

\subsection{Major QAs}
\label{sec:determine-impact-on-qa}
Quality attributes (QA) may be described as the factors that have a system-wide impact on an application's architecture, implementation, and operation \cite{bass-sap-3ed-2012}. The QAs of concern for most of applications may be categorized depending on whether they affect design-, runtime-, system- or user- qualities of the application  \cite{swebokv3,ms-arch-guide}. 

The overall design and quality of many software applications can be considered acceptable when the applications possess a reasonable level of at least the following QAs  \cite{bass-sap-3ed-2012,bass2000quality,qa-models-ISO-IEC}: Performance, Reliability, Scalability, Security, and Usability. However, we have considered a slightly expanded list of QAs when examining how they are affected for a software which is built for using a QCP:


\begin{multicols}{2}
\begin{enumerate}
    \item Availability
    \item Interoperability
    \item Maintainability
    \item Manageability
    \item Performance
    \item Reliability
    \item Scalability
    \item Security 
    \item Testability
    \item Usability
\end{enumerate}
\end{multicols}

The current body of software design and architecture knowledge contains extensive detail and discussions about:
\begin{itemize}
\item The definitions of various QAs. E.g.  \cite{swebokv3,bass-sap-3ed-2012,ms-arch-guide,qa-models-ISO-IEC} are significant references.
\item How to realize a QA in software using various design tactics. E.g.  \cite[Part-II]{bass-sap-3ed-2012} \cite{bass2000quality}.
\item What characteristics of a computing platform are leveraged by various tactics in order to realize a QA. E.g.  \cite{bachmann2002illuminating,deriving-tactics,sodhi-assess-platform-qa-2012,Sodhi:2011-assess-cloud}.
\end{itemize}

Thus, we have not covered such details here. However, when identifying the impact of various QCP characteristics on different QAs, we have relied mainly on that knowledge body. In the subsequent paragraphs, we discuss only those aspects of a QA that are relevant for identifying how the QA is affected by various characteristics of QCPs. We take each of the QCP characteristics identified in Section \ref{sec:qcp-char-impact} in turn and discuss how it affects various QAs under consideration. A characteristic may have a \textit{favorable, unfavorable} or, \textit{neutral} impact for building a QA into an application.

\subsection{SDLC activities}
\label{sec:determine-impact-on-sdlc}
Core activities involved in software development are \cite{swebokv3}: Requirements, Design Engineering, Construction, Testing, Debugging, Deployment, and Maintenance. 
We considered the following to determine which of these core activities are likely to be affected when developing software for QCPs: 
\begin{itemize}
\item The definitions and properties of various SDLC activities, as defined in SWEBOK  \cite[\S 2 of Ch. 8]{swebokv3}.
\item Mapping of various QAs to different SDLC activities. The mapping is done based on the relevance of an SDLC activity for the realization of a QA. Table \ref{tab:qa-sdlc-mapping} shows the mapping that has been derived using the definitions and properties of QAs  \cite[\S 5 of Ch. 2]{swebokv3} and SDLC activities. Once we know whether QCP characteristics impact a QA, we can use this mapping to cross-check whether an SDLC activity is likely to be affected.
\end{itemize}

\begin{table}[!h]
\caption{Mapping$^*$ of QAs to SDLC activities}
\label{tab:qa-sdlc-mapping}
\begin{center}
\resizebox{0.8\linewidth}{!}{
\begin{tabular}{c|c|c|c|c|c|c}


\multirow{2}{*}{\shortstack{Quality\\ Attributes}}& \multicolumn{6}{c}{SDLC Activity}\\
\cline{2-7}
& \rotatebox{90}{Requirements} & \rotatebox{90}{Design} & \rotatebox{90}{Construction} & \rotatebox{90}{Testing} & \rotatebox{90}{Deployment} & \rotatebox{90}{Maintenance} \\
\toprule
Availability & L & H & H & L & H & L \\  
Interoperability & L & H & H & L & L & L \\  
Maintainability & L & H & H & L & L & H \\  
Manageability & L & H & H & L & H & L \\  
Performance & L & H & H & L & H & L \\  
Reliability & L & H & H & L & H & L \\  
Scalability & L & H & H & L & H & L \\  
Security  & L & H & H & L & H & L \\  
Testability & L & H & H & H & L & L \\  
Usability & L & H & H & L & L & L \\  
\bottomrule
\multicolumn{7}{l}{$^*$Cell value indicates level of relevance: (H)igh, (L)ow}
\end{tabular}
}
\end{center}
\end{table}

\section{Discussion of QCP characteristics and their impact}
\label{sec:qcp-char-impact}
In this section, we describe the critical characteristics of  QCPs (some discussed in Section \ref{sec:implementation details}) and the impact of each characteristic on various QAs and SDLC activities. 
The details about the impact on QAs have been presented in-line with the description of a characteristic. Further, the discussion about the impact on SDLC activities has been presented in a dedicated section (Section \ref{sec:effect-swp}). We hope such a presentation will make it easier to understand (and convey) those details without loss of relevant information. 

\subsection{Major QCP characteristics and their impact on QAs}
\label{sec:qcp-chars-and-qa-impact}
\subsubsection{Lower level of the programming abstractions}
The programming abstractions offered by QCPs of today \cite{ibm-qx,dwave-tech-overview,ms-quantum-sdk,open-qasm} are of a low level. That is, a programmer must work at the level of quantum logic gates (usually via a high-level language representation of it)  when expressing computational steps that he wants to execute via a QPU. 

\textbf{Effect on QAs:} Working at the quantum gates level to implement a quantum algorithm is complicated. To develop programs using such low levels of abstraction is error-prone  \cite{sw-maint-economics-2006} and increases the complexity of the code. Moreover, there are not many (as of today at least) expert quantum programmers, which adversely affect QAs such as maintainability, testability, reliability, and availability \cite{bachmann2002illuminating}.
    
\subsubsection{Platform heterogeneity} 
The published technical details of different QCPs  \cite{dwave-tech-overview,ibm-qx,ms-quantum,rigetti-qvm} that are available today show the heterogeneous nature of the QCPs. 
Furthermore, various quantum no-go theorems (please see Section \ref{sec:key-quantum-principles}) disallow copy, deletion, and qubit states' transmission under certain conditions. Due to restrictions such as these, the dependency on classical hardware or software is inevitable. Thus, the QCPs are expected to be inherently \textit{heterogeneous}. That is, both classical and quantum hardware and software are involved.

\textbf{Effect on QAs:} Heterogeneity makes it challenging to implement high software-cohesion  \cite[Part-II]{bass-sap-3ed-2012,bass2000quality}. Therefore, QAs such as maintainability, reliability, robustness, reusability, and understandability get adversely affected  \cite{bachmann2002illuminating,deriving-tactics} due to low cohesion. The heterogeneous environment also means more  disparate elements (software and hardware) to be managed, thus adversely affecting the manageability and testability of the system.
    
\subsubsection{Physical environment}
Quantum hardware requires an entirely different type of physical environment. Most of the existing implementations \cite{dwave-hardware,dwave-tech-overview,ibm-qx,ms-quantum} of quantum hardware circuitry use superconducting loops requiring ultra-low temperatures. Achieving and maintaining such physical conditions necessitates a very specialized environment for the reliable operation of a QCP.

\textbf{Effect on QAs:} A highly specialized physical environment needed by a QCP is challenging to create, maintain, and operate. The effect of ambient noise and interference is more pronounced in the case of QPUs than it is on CPUs. Such properties adversely affect the QAs such as availability (e.g., due to \enquote{brittle} nature of physical qubits), manageability, scalability (e.g., due to difficulties in quickly launching additional QPU instances), and testability of the system  \cite{bachmann2002illuminating,deriving-tactics,bass-sap-3ed-2012}.

\subsubsection{Large form factor}  
Due to the requirement of a particular physical environment, the quantum computers of today are massive. For instance, the main box of D-Wave\texttrademark 2000Q system measures   roughly $10ft*10ft*10ft$ in size \cite{dwave-hardware}.

\textbf{Effect on QAs:} Most of the QAs, except scalability, remain unaffected by this property of QCPs. The large form factor makes it difficult to augment the QCP capabilities, thus adversely affecting the scalability \cite{bachmann2002illuminating,deriving-tactics,bass2000quality}.

\subsubsection{Energy Efficiency} 
Energy consumption can be looked at from two main aspects: energy consumed by the QPU, the energy requirements for cooling, and other ancillary circuitry make up a quantum computer. It has been observed that a quantum computer spends  \cite{quantum-energy-consumption} most of its energy on cooling. A QPU by itself, however, consumes much less energy  \cite{energy-efficient-qcp-2017}. As the computation speeds grow to exascale (i.e., 1000 petaflops), the energy consumption of a QPU is not expected to increase as fast as that of a CPU or a GPU. Experiments conducted with D-Wave's 2000-qubit system showed  \cite{dwave-energy} overall energy efficiency improvements of the order of 100x compared to state-of-the-art classical computing servers when considering pure computation time for running specialized algorithms.

\textbf{Effect on QAs:} Except for improving performance and scalability, this property does not significantly impact most other QAs in our list.

\subsubsection{Remote software development and deployment}
Programming tools such as IDEs, debuggers, and simulators. that a software developer can use to create quantum software are invariably cloud-based  \cite{ibm-qx,dwave-tech-overview}. Only a minimal portion  \cite{ms-quantum-sdk,open-qasm} of the quantum programming tools stack can be deployed and used on a programmer's local machine. For the development and testing of production-ready quantum software, a programmer typically requires access to a remote QCP environment.

\textbf{Effect on QAs:} The development tools and environment used for developing software for (or on) a QCP are usually distributed in nature. The decentralized (typically, offered via a cloud-based IDE) and the development environment's heterogeneous nature make programming, testing, and debugging quantum programs slower and tedious \cite{cloud-based-development}, which adversely affects the maintainability and the testability \cite{bachmann2002illuminating,deriving-tactics,bass2000quality}.

\subsubsection{Dependency on the Known quantum algorithms}
To exploit a quantum computer's real potential, a programmer must express their software logic using quantum algorithms.  A computing task where one is looking to gain speed-up by running it on a QCP is typically mapped to or broken into another task(s) for which a quantum algorithm(s) is known.

\textbf{Effect on QAs:}  
There are very few  \cite{quantum_algo_progress_2004} -- only about a dozen -- quantum algorithms known for different types of problems. Software engineers have to map their problems to one of the existing few quantum algorithms, which adversely affects the ability to perform enhancement and corrective maintenance.

Moreover, the probabilistic nature of quantum computations and their results adversely affect testability and interoperability (with classical software). Synthesizing realistic pairs of $\langle input, expected~output \rangle$ for test case scenarios and reproducing the defects that one wants to debug requires particular approaches.

\subsubsection{Limited portability of software}
Quantum computing platforms are themselves in their infancy, though under rapid growth. There is a lack of standards necessary for developing quantum programs that can be executed transparently on different QPUs. Each of the significant providers of QCPs offers their proprietary programming APIs and tools   \cite{dwave-tech-overview,ms-quantum-sdk,rigetti-quantum-isa-2016,rigetti-qvm,ibm-qx,open-qasm}. 

\textbf{Effect on QAs:} QCPs lack standardization in several areas ranging from high-level programming APIs to low-level hardware. For example, a quantum program written using Rigetti's quantum ISA  \cite{rigetti-quantum-isa-2016} may not be executable on Open QASM \cite{open-qasm} supported by IBM. The portability of software is adversely affected by QCPs. Lack of portability adversely affects  \cite[Part-2]{bass-sap-3ed-2012}  \cite{bass2000quality} availability, interoperability, maintainability, and scalability.

\subsubsection{Limited quantum networking}
Though long-distance quantum entanglement distribution is feasible \cite{satellite-based-entanglement-distrib-2017}, realizing practical quantum communication networks is still a work in progress \cite{towards-quantum-network-simon_2017,quantum-network-kimble-2008}. Moreover, various no-go theorems (please see Section \ref{sec:key-quantum-principles}) pose further restrictions on extracting qubit states. Practical quantum software that depends on the availability of a reliable quantum network would be hard to achieve.

\textbf{Effect on QAs:} A reliable quantum network is necessary for building reliable and high-performance quantum software, which make the development of a non-trivial quantum software tedious and error-prone. Quantum networks are not yet production-ready  \cite{towards-quantum-network-simon_2017,quantum-network-kimble-2008}. Thus, the performance, scalability, reliability, and availability will be adversely affected.

\subsubsection{Lack of native quantum operating system}
\label{sec:impact-qa-native-os}
The quantum processors are still controlled via classical computing operating systems  \cite{quantum-os}. We do not yet have mature multitasking and multiprocessing capabilities available for quantum processors. Most existing QCPs  \cite{ibm-qx,dwave-tech-overview,ms-quantum-sdk,ms-quantum} expose the quantum gates and qubits for direct manipulation by programmers. Mature protocols and APIs that implement, for example, practical timesharing of a QPU are not available yet  \cite{quantum-os}. Similarly, quantum algorithms that may exploit multiple QPUs in parallel are yet to be explored.

\textbf{Effect on QAs:} A native operating system helps harness the full potential of hardware securely and effectively, which is lacking  \cite{quantum-os} in the case of QCPs. It prohibits, for instance, practical timesharing of a QPU. Thus, the performance, manageability, reliability, scalability, and security will be adversely affected  \cite[Part-2]{bass-sap-3ed-2012}  \cite{bass2000quality}.
    
\subsubsection{Limited multitasking and multiprocessing}
These result due to the lack of native quantum operating systems.  A programmer must rely on classical computer's OS for achieving any type of multitasking and multiprocessing on a given set of QPUs. 

\textbf{Effect on QAs:} This scenario is the same as the QA effect described in Section \ref{sec:impact-qa-native-os}.
    
\subsubsection{Fundamentally different programming model} 
As discussed in Section \ref{sec:q-software-dev}, quantum programs are inherently probabilistic. A programmer looking to harness the power of a QCP must identify or design suitable quantum algorithms that can solve the problem at hand. 

\textbf{Effect on QAs:} Quantum programs require a fundamentally different approach to programming (please see Section \ref{sec:q-programming-model} ), which affects the ease of use of the underlying technology and, in turn, the code complexity. Both of these are essential factors that influence  \cite{sw-maint-economics-2006} the development and maintenance of dependable software. 
Thus, a QCP adversely affects the maintainability, interoperability, security, and testability QAs.

\subsubsection{Dependency on classical storage} 
Though limited time storage of entangled qubits is feasible  \cite{quantum-storage-2011,quantum-storage-2014}, long term persistence of qubits in passive media is still impossible. Besides, it is not feasible to store arbitrary non-entangled qubits due to different no-go theorems such as no-teleportation  \cite[Page 128]{pathak2013elements}, no-cloning  \cite{no-cloning-wootters1982}, and no-deleting  \cite{no-deleting-pati2000} theorems. The quantum no-go theorems disallow copy, deletion, the transmission of qubit states under certain conditions. Thus, the permanent persistence of critical data in quantum programs still requires classical storage devices. 

\textbf{Effect on QAs:} For durably persisting critical data, the quantum programs depend on classical storage devices due to different no-go theorems (please see Section \ref{sec:key-quantum-principles}). The quantum no-go theorems disallow copying, deletion, or transmission of qubits under certain conditions. Such restrictions on a QCP adversely affect the manageability, performance, and scalability.

\begin{table}[!t]
		\centering
		\caption{Impact$^*$ of QCP characteristics on QAs\label{tab:qa-impact}}
		\resizebox{\linewidth}{!}
		{
		\begin{tabular}{l *{10}{|c}}	
		
		\makecell{\textbf{QAs $\rightarrow$} \\ \textbf{Characteristics$\downarrow$}} & 
		\rotatebox{90}{Availability} & 
		\rotatebox{90}{Interoperability} &
		\rotatebox{90}{Maintainability} &
		\rotatebox{90}{Manageability} &
		\rotatebox{90}{Performance} &
		\rotatebox{90}{Reliability} &
		\rotatebox{90}{Scalability} &
		\rotatebox{90}{Security} &
		\rotatebox{90}{Testability} &
		\rotatebox{90}{Usability} \\
		
		\toprule
		Platform heterogeneity & 
		U & -- & U & U & -- & U & -- & -- & U & -- \\
		
		Special physical environment & 
		U & -- & -- & U & -- & U & U & U & U & -- \\
		Large form factor & 
		-- & -- & -- & -- & -- & -- & U & -- & -- & -- \\
		Higher energy efficiency & 
		-- & -- & -- & -- & F & -- & F & -- & -- & -- \\
		\multirow{2}{*}{\shortstack{Lower level of the-\\ programming abstractions}} &
		\multirow{2}{*}{U} & \multirow{2}{*}{U} & \multirow{2}{*}{U} & \multirow{2}{*}{--} & \multirow{2}{*}{--} & \multirow{2}{*}{U} & \multirow{2}{*}{--} & \multirow{2}{*}{--} & \multirow{2}{*}{U} & \multirow{2}{*}{--} \\
		& & & & & & & & & & \\
		\multirow{2}{*}{\shortstack{Remote software development\\ and deployment}} &
			\multirow{2}{*}{--} & 	\multirow{2}{*}{--} & 	\multirow{2}{*}{U} & 	\multirow{2}{*}{--} & 	\multirow{2}{*}{--} & 
				\multirow{2}{*}{--} & 	\multirow{2}{*}{--} & 	\multirow{2}{*}{--} & 	\multirow{2}{*}{U} & 	\multirow{2}{*}{--} \\
				& & & & & & & & & & \\
		\multirow{2}{*}{\shortstack{Dependency on quantum\\ algorithms}} & 
		\multirow{2}{*}{--} & \multirow{2}{*}{U} & \multirow{2}{*}{U} & \multirow{2}{*}{--} & \multirow{2}{*}{F} & \multirow{2}{*}{--} & \multirow{2}{*}{F} & \multirow{2}{*}{--} & \multirow{2}{*}{U} & \multirow{2}{*}{--} \\
				& & & & & & & & & & \\
		\multirow{2}{*}{\shortstack{Limited portability\\ of software}} &
		\multirow{2}{*}{U} & \multirow{2}{*}{U} & \multirow{2}{*}{U} & \multirow{2}{*}{--} & \multirow{2}{*}{--} & \multirow{2}{*}{--} & \multirow{2}{*}{U} & \multirow{2}{*}{--} & \multirow{2}{*}{--} & \multirow{2}{*}{--} \\
		& & & & & & & & & & \\
		\multirow{2}{*}{\shortstack{Limited quantum\\ networking}} & 
		\multirow{2}{*}{U} & \multirow{2}{*}{--} & \multirow{2}{*}{--} & \multirow{2}{*}{--} & \multirow{2}{*}{U} & \multirow{2}{*}{U} & \multirow{2}{*}{U} & \multirow{2}{*}{--} & \multirow{2}{*}{--} & \multirow{2}{*}{--} \\
		& & & & & & & & & & \\
		\multirow{2}{*}{\shortstack{Lack of native quantum-\\ operating system}} &
		\multirow{2}{*}{--} & \multirow{2}{*}{--} & \multirow{2}{*}{--} & \multirow{2}{*}{U} & \multirow{2}{*}{U} & \multirow{2}{*}{U} & \multirow{2}{*}{U} & \multirow{2}{*}{U} & \multirow{2}{*}{--} & \multirow{2}{*}{--} \\
		& & & & & & & & & & \\
		\multirow{2}{*}{\shortstack{Fundamentally different\\ programming model}} &
		\multirow{2}{*}{--} & \multirow{2}{*}{U} & \multirow{2}{*}{U} & \multirow{2}{*}{--} & \multirow{2}{*}{--} & \multirow{2}{*}{U} & \multirow{2}{*}{--} & \multirow{2}{*}{U} & \multirow{2}{*}{U} & \multirow{2}{*}{--} \\
		& & & & & & & & & & \\
		\multirow{2}{*}{\shortstack{Dependency on\\ classical storage}} &
		\multirow{2}{*}{--} & \multirow{2}{*}{--} & \multirow{2}{*}{--} & \multirow{2}{*}{U} & \multirow{2}{*}{U} & \multirow{2}{*}{U} & \multirow{2}{*}{U} & \multirow{2}{*}{--} & \multirow{2}{*}{--} & \multirow{2}{*}{--} \\
		& & & & & & & & & & \\
		\bottomrule
		\multicolumn{11}{l}{$^*$Cell value indicates impact on QA: (F)avorable, (U)nfavorable, \enquote{--} $\Rightarrow$ Unknown, Neutral}
		\end{tabular}	
		}
\end{table}

\begin{table}[!th]
		\centering
		\caption{Impact$^\dagger$ of QCP characteristics on SDLC activities\label{tab:sdlc-impact}}
		\resizebox{0.8\linewidth}{!}
		{
		\begin{tabular}{l *{6}{|c}}
		
		
		\makecell{\textbf{SDLC Activity $\rightarrow$} \\ \textbf{Characteristics$\downarrow$}}  & 
		\rotatebox{90}{Requirements} & 
		\rotatebox{90}{Design} &
		\rotatebox{90}{Construction} &
		\rotatebox{90}{Testing} &
		\rotatebox{90}{Deployment} &
		\rotatebox{90}{Maintenance} \\
		
		\toprule
		
		Platform heterogeneity & Y & Y & Y & Y & Y & Y \\ 
		& & & & & &\\
		Special physical environment & N & Y & Y & Y & Y & Y \\ 
		& & & & & & \\
		Large form factor & N & Y & N & Y & N & Y \\ 
		& & & & & &\\
		Higher energy efficiency & N & Y & N & N & N & N \\ 
		& & & & & &\\
		\multirow{2}{*}{\shortstack{Lower level of the\\ programming abstractions}} & \multirow{2}{*}{N} & \multirow{2}{*}{Y} & \multirow{2}{*}{Y} & \multirow{2}{*}{Y} & \multirow{2}{*}{N} & \multirow{2}{*}{Y} \\
	& & & & & &\\
	& & & & & &\\
		\multirow{2}{*}{\shortstack{Remote software development\\ and deployment}} & \multirow{2}{*}{N} & \multirow{2}{*}{Y} & \multirow{2}{*}{Y} & \multirow{2}{*}{Y} & \multirow{2}{*}{Y} & \multirow{2}{*}{Y} \\
		& & & & & &\\
		& & & & & &\\
		\multirow{2}{*}{\shortstack{Dependency on quantum\\ algorithms}} & \multirow{2}{*}{N} & \multirow{2}{*}{Y} & \multirow{2}{*}{Y} & \multirow{2}{*}{Y} & \multirow{2}{*}{N} & \multirow{2}{*}{Y} \\
		& & & & & &\\
		& & & & & &\\
		\multirow{2}{*}{\shortstack{Limited portability\\ of software}} & \multirow{2}{*}{N} & \multirow{2}{*}{Y} & \multirow{2}{*}{Y} & \multirow{2}{*}{N} & \multirow{2}{*}{Y} & \multirow{2}{*}{N} \\ 
		& & & & & &\\
		& & & & & &\\
		\multirow{2}{*}{\shortstack{Limited quantum\\ networking}} & \multirow{2}{*}{N} & \multirow{2}{*}{Y} & \multirow{2}{*}{Y} & \multirow{2}{*}{N} & \multirow{2}{*}{N} & \multirow{2}{*}{N} \\
		& & & & & &\\
		& & & & & &\\
		\multirow{2}{*}{\shortstack{Lack of native quantum-\\ operating system}} & \multirow{2}{*}{N} & \multirow{2}{*}{Y} & \multirow{2}{*}{Y} & \multirow{2}{*}{N} & \multirow{2}{*}{Y} & \multirow{2}{*}{N} \\
		& & & & & &\\
		& & & & & &\\
		\multirow{2}{*}{\shortstack{Fundamentally different\\ programming model}} & \multirow{2}{*}{N} & \multirow{2}{*}{Y} & \multirow{2}{*}{Y} & \multirow{2}{*}{Y} & \multirow{2}{*}{N} & \multirow{2}{*}{Y} \\
		& & & & & &\\
		& & & & & &\\
		
		\multirow{2}{*}{\shortstack{Dependency on classical-\\ storage}} & \multirow{2}{*}{N} & \multirow{2}{*}{Y} & \multirow{2}{*}{Y} & \multirow{2}{*}{Y} & \multirow{2}{*}{N} & \multirow{2}{*}{N} \\ 
		& & & & & &\\
		
		\bottomrule
		\multicolumn{7}{l}{$\dagger$Cell values indicate whether SDLC activity is impacted: (Y)es, (N)eutral}
		\end{tabular}}
		
\end{table}

\subsection{Effect on the SDLC activities}
\label{sec:effect-swp}
Core activities involved in software development are \cite{swebokv3}: Requirements, Design Engineering, Construction, Testing, Debugging, Deployment, and Maintenance. We examine how each of these core activities is likely to be affected when building software for QCPs.

We considered the standard definitions of various SDLC activities as provided in SWEBOK  \cite{swebokv3} for analyzing how those activities may be affected by the characteristics of quantum platforms.

\subsubsection{Requirements engineering} Considering the definition of requirements engineering tasks as outlined in SWEBOK  \cite{swebokv3}, one can safely deduce that, except for the analysis and validation, most requirements engineering tasks in themselves are not affected  \cite[Ch. 1]{swebokv3} by the nature of a target computing platform. An essential and commonly employed tool for requirements validation  \cite[\S 4.3 of Ch. 1]{swebokv3} is \textit{prototyping}. Developing a prototype demands hands-on knowledge of the target platform \cite[Ch. 7]{sommerville2007software}  \cite{ms-arch-guide}.

The software architectural characteristics of QCPs will affect the analysis and allocation of requirements and the requirements validation via prototyping. For example, lack of ability to store or copy the exact state of qubits to classical bits will make it impossible to realize use cases which depend on such copying of information from the QCP. Similarly, the simulators that are available for QCPs will pose difficulties for reliable prototyping because those simulators can represent QPUs having a very small number of qubits.
    
\subsubsection{Design engineering} Software design is a critical activity in software development, as its primary goal is to deliver models that define the blueprint of the solution to be implemented. Design mainly deals with two significant aspects of the software development a) defining the structural (de)composition of the system, and b) addressing essential quality attributes such as scalability, performance, reliability, security, and usability \cite[Ch. 2]{swebokv3}  \cite[Ch. 7]{sommerville2007software}.

Almost all the QCP characteristics listed in Section \ref{sec:qcp-char-impact} affect the software's design, which is to be developed for a QCP. For instance, centralized deployment of quantum programs would limit the design choices a software architect has when distributing components  \cite{bass-sap-3ed-2012,richards2015software} of a solution. QCP characteristics such as dependency on quantum algorithms, and the limited portability of quantum programs, affect the ability to realize \cite{bachmann2002illuminating,deriving-tactics,bass2000quality} various QA's such as portability, testability, and maintainability of the quantum software.
Moreover, native quantum OS services related to multitasking and multiprocessing on a QCP are not yet available. This further limits the design choices for building scalability into the quantum software.

\subsubsection{Construction} This SDLC phase deals with the working software product as per the design arrived during the preceding phase. The effects of having adopted a QCP for implementing parts of the software are most prominent in this phase. There are three prominent aspects of QCPs that affect this phase (and also few others): i) Programming model, ii) Heterogeneous nature of a QCP, and iii) Decentralized development environment.

Because quantum computing takes a fundamentally different approach to programming (i.e., expressing the logic or algorithm), algorithms are invariably probabilistic here. A significant implication of QCPs concerning software construction is the lack of a rich set of \textit{quantum software libraries}. As of today, there are about twenty quantum algorithms known for different types of problems \cite{coles2018quantum}. Software engineers have to map their problems to one of the existing quantum algorithms. The most critical limitation of QCPs during this phase is that they only have about a dozen quantum algorithms, and not every problem can be mapped easily to one of these algorithms.

Another implication of writing software for a QCP is that for any serious programming, one requires access to at least a reasonably powerful machine that can simulate a QCP. A typical PC alone does not suffice. Moreover, access to a real quantum computer may be a hard requirement when writing programs for certain types of application scenarios. Therefore, developing software for a QCP is expected to follow a \textit{client-server} model where the programmers use their PCs to program remotely (perhaps via some cloud-based IDE) on a QCP. Programming on a remote machine or an IDE is known  \cite{cloud-based-development} to limit a programmer's productivity.
    
\subsubsection{Testing and debugging} These activities are significantly impacted by i) ability to detect and isolate faults in the software, ii) ease of creating test assessment criteria for the system and its components, and iii) the ease of executing these tests for checking if the criteria are met. A significant implication for these activities arises from the probabilistic nature of quantum computations and their results.  Synthesizing realistic pairs of $\langle input, expected~output \rangle$ for test case scenarios and reproducing the defects that one wants to debug will require particular approaches. 

Further, as highlighted in the \textit{Construction} case, the decentralized (typically, offered via a cloud-based IDE) and heterogeneous nature of the development environment will make testing and debugging of quantum programs slower and tedious  \cite{cloud-based-development}. It mainly happens because a QCP introduces additional layers (please see Fig. \ref{fig:arch}) of communication.
    
\subsubsection{Deployment} These activities are affected by factors such as i) The ease with which system administrators can manage the software. Typically, one relies on the instrumentation embedded in the system for allowing system monitoring, debugging, and performance tuning. ii) Availability of mature tooling support for the ease of making changes to a system.
Two significant aspects of QCPs from a software deployment perspective are a) The heterogeneous nature of the overall system (please see Fig. \ref{fig:arch}), and b) The (un)availability of full-capability QCPs to an organization or country.

As has been already highlighted, the quantum software applications will have a distributed deployment model comprising classical, and quantum software components of that application, which is likely to remain the case until entirely native QCPs with native I/O and operating systems become a reality.  Such heterogeneity is inevitable because i) measuring the state of a qubit forces it to collapse to a classical bit state, and ii) difficulties due to no-go theorems  in copying and moving data directly between qubits and other classical devices \cite{no-cloning-wootters1982,no-deleting-pati2000}.

Direct access to the full-power QCPs may be difficult, if not impossible, for many organizations due to the potential for misuse, the complexity, and sophistication of the QCPs. Such a scenario implies that the access to QCPs will be controlled, and interfacing will happen over the network. Thus, requiring a distributed deployment model for software applications.
    
\subsubsection{Maintenance}
Maintenance of software can be classified  \cite{sw-maintenance} as either a \textit{Correction} or an \textit{Enhancement}. The former can be of two types, viz., corrective or preventive, while the latter can be adaptive or perfective. Regardless of the type of maintenance, the following are a few of the essential factors that influence  \cite{sw-maint-economics-2006} maintenance of software: staff experience, code complexity, level-programming languages or abstractions, and the ease-of-use of the underlying technology.

Considering their characteristics, the QCPs of today score poorly on each of these factors. For instance, programming the QCPs requires one to work at the quantum gates level \cite{dwave-tech-overview,ibm-qx,ms-quantum-sdk} to compose the necessary quantum circuits to implement a quantum algorithm. Working at such low-levels of abstraction is not easy and increases the complexity of the code. Moreover, there are not many (as of today at least) expert quantum programmers. Thus, maintenance (and development in general) of software is not easy on QCPs.

\subsection{Applications of QA impact information}
\label{sec:applications}
One of the critical outcomes that we have presented in this paper is the information about how the QCPs impact various QAs and SDLC activities. This information's significant applications can be in the evaluation of design decisions and computing platform selection. The information presented in Table \ref{tab:qa-impact} and Table \ref{tab:sdlc-impact} can be easily used by a design decision support tool or method, such as the one presented by  \cite{sodhi-assess-platform-qa-2012}. Typically, such tools make use of Multi-Criteria Decision Making techniques (e.g., MAUT  \cite{MAUT:Wallenius:2008}, and TOPSIS  \cite{sodhi2012simplified}) to evaluate a \enquote{choice} by considering the impact of the contributing factors on that \enquote{choice.} The \enquote{choice} here could be a design decision, a computing platform such as QCP vs. non-QCP.

\subsection{Threats to validity}
\label{sec:threats2v}
Quantum computing is a fast-evolving area of technology. The characteristics of QCPs that we have identified are based on the study of currently available state-of-the-art quantum hardware and software. We expect advances in quantum computing will invalidate a few of these characteristics in the coming years. For example, production-ready native quantum operating systems are likely to be feasible in the years to come. 

Further, the list of characteristics that we have given is not exhaustive. It is likely that additional QCP characteristics, taken together, may be of significance in specific software development scenarios. We would like to highlight that our findings are derived from: a) published information about QCPs and b) experimental programming on the QCPs accessible to us. There are likely unpublished features or information about those QCPs that can affect quantum software applications' software architecture.  Next, we have not covered every QA, which is relevant for a broader set of application types. There may be QAs such as auditability, distributability, and extensibility, which are relevant and may be affected by the QCPs.

\section{Conclusions}
Programmers' interest in applying quantum computing has surged in the recent past. Leveraging this technology in solving scientific or business problems requires a deeper understanding of essential characteristics of QCPs, particularly those relevant for software development. 

We have shown that the critical characteristics of a QCP make it entirely different from a classical computing platform. For instance, availability and know-how of quantum algorithm(s) for solving a task at hand are a hard requirement for developing quantum software applications. Such characteristics of a QCP affect various QAs of the quantum software application. The QCP characteristics such as limited portability of quantum programs, inability to copy, move, or delete qubits data under certain conditions, and the dependency on quantum algorithms, adversely affect the ability to realize various QAs such as scalability, portability, testability, and maintainability of the quantum application software. The QAs that are favorably impacted by QCP characteristics include performance and scalability. The study of QCP's impact on QAs can be beneficial for design decision evaluation tasks. Further, most of the SDLC activities require special handling when building software for a QCP.

Overall, the specialized nature of QCPs appears to have a significant implication: it limits the use of QCPs for very specialized application areas where the \textit{performance} QA is of chief importance. However, because quantum computing is undergoing rapid development, we expect that this technology's evolution will likely introduce additional factors that with high likelihood will affect the architecture of quantum software applications.

\bibliographystyle{IEEEtran}
\bibliography{IEEEabrv,bibFile}
\end{document}